\documentclass[aps, prb, twocolumn, superscriptaddress, footinbib, preprintnumbers, footinbib, 10pt]{revtex4-2}

\usepackage{ae,aecompl} 
\usepackage{float}
\usepackage{graphicx}
\usepackage{dcolumn}
\usepackage{amsmath, mathrsfs}
\usepackage{amsfonts}
\usepackage{amsthm}
\usepackage{bm}
\usepackage[colorlinks=true, linkcolor=red, citecolor=blue, urlcolor=magenta]{hyperref}
\usepackage{orcidlink}

\newcommand{\orcidBhargava}{\orcidlink{0009-0006-7595-8776}}
\newcommand{\orcidKamlesh}{\orcidlink{0009-0002-5976-0234}}
\newcommand{\orcidGouranga}{\orcidlink{0009-0006-4339-1373}}
\newcommand{\orcidAniruddha}{\orcidlink{0009-0006-3178-3732}}
\newcommand{\orcidVijay}{\orcidlink{0000-0002-6849-0451}}
\newcommand{\orcidDebarag}{\orcidlink{0000-0002-8195-9920}}
\newcommand{\orcidArun}{\orcidlink{0000-0003-4844-0155}}

\begin{document}

\title{Loan Portfolio Optimization with Variational Quantum Algorithms}

\author{Balaganchi~A.~Bhargava\orcidBhargava}
\affiliation{LTM Research, LTM}
 
\author{Kamlesh~Kumar\orcidKamlesh}
\affiliation{LTF Lab, L\textnormal{\&}T Finance}

\author{Gouranga~Dinda\orcidGouranga}
\affiliation{LTM Research, LTM}

\author{Aniruddha~Biswas\orcidAniruddha}
\affiliation{LTM Research, LTM}

\author{Vijay~S.~Rao\orcidVijay}
\affiliation{LTM Research, LTM}

\author{Debarag~Banerjee\orcidDebarag}
\affiliation{LTF Lab, L\textnormal{\&}T Finance}

\author{Arun~Sehrawat\orcidArun}
\email{arun@ltm.com}
\affiliation{LTM Research, LTM}

\date{September 24, 2026}

\begin{abstract}
Loan portfolio optimization (LPO) seeks portfolios that minimize credit risk while satisfying practical selection constraints. Accounting for portfolio-level risk requires modeling not only expected losses but also loss variability and borrower-default correlations, leading to a large-scale combinatorial optimization problem whose complexity grows rapidly with portfolio size. In this work, we formulate LPO as a Quadratic Constrained Binary Optimization (QCBO) problem that jointly captures expected and unexpected losses and transform the resulting formulation into a Quadratic Unconstrained Binary Optimization (QUBO) model suitable for variational quantum optimization. To address the limited qubit resources of current quantum hardware, we employ Pauli Correlation Encoding (PCE), which enables a large number of portfolio-selection variables to be represented using comparatively few qubits. The proposed framework is evaluated on a real-world microfinance dataset comprising $2012$ borrowers. For problem instances with up to $1000$ portfolio-selection variables, we benchmark classical simulations of a variational quantum algorithm (VQA) against Google OR-Tools under matched time budgets, obtaining feasible portfolios with an objective-value gap of about $40\%$ across all sizes. We further implement the framework on superconducting quantum hardware for instances containing up to 1500 variables. Across these instances, the experimentally obtained objective values are within $10\%$ of the corresponding simulation results. These findings demonstrate the feasibility of applying PCE-based variational quantum optimization to large-scale loan portfolio optimization under current quantum hardware limitations.
\end{abstract}

\maketitle

%=========================================================
\section{Introduction}\label{sec:Intro}

Debt portfolios differ fundamentally from equity portfolios in several respects. Debt instruments, such as loans and bonds, generally have predefined contractual cash flows, whereas equities offer uncapped upside potential. Equity prices are updated continuously in the market, whereas loan portfolio data are typically refreshed much less frequently. Moreover, both equity returns and credit losses may exhibit non-normal behavior; however, loan portfolios are characterized by default-driven, skewed, and fat-tailed loss distributions that require distinct modeling approaches~\cite{JPMcreditmetrics1997, creditriskplus1997, bluhm2010}.

Loans constitute a major asset class for banks and non-banking financial companies (NBFCs), making credit-risk management central to lending operations. 
To support credit decisions, financial institutions increasingly employ statistical and machine-learning models that leverage borrower characteristics, financial history, and macroeconomic information to estimate risk metrics such as the probability of default (PD)~\cite{louzada2016, thomas2017, hand1997, baesens2003, lessmann2015, RolandAbi2025, Bellotti2009}. 
While these models improve individual borrower assessment, lending portfolio risk depends on collective effects that cannot be inferred from borrower-level predictions alone. 
In particular, exposures to common economic, geographic, sectoral, or behavioral factors can induce \emph{correlated defaults}, resulting in portfolio losses that differ substantially from those implied by independent risk estimates~\cite{gordy2003, frey2001, bluhm2010}. Furthermore, borrower-level default probabilities primarily contribute to expected-loss estimates and do not fully characterize the \emph{variability of losses}~\cite{bluhm2010, frey2001}. 
Consequently, portfolio construction requires optimization frameworks that account for expected losses, loss variability, and correlations among borrowers.

Loan portfolio optimization (LPO) addresses this challenge by selecting borrowers from a candidate pool while satisfying business, operational, and regulatory constraints. 
Beyond minimizing expected credit losses, LPO must control concentration and correlation risks that can amplify losses under adverse conditions~\cite{andersson2001, wang2022, okawara2023, frey2001, bluhm2010}. 
This naturally leads to portfolio-level optimization models that jointly consider expected and unexpected losses~\cite{JPMcreditmetrics1997, bluhm2010, andersson2001}. 
As the number of borrowers and constraints increases, LPO becomes a challenging combinatorial optimization problem.

The computational difficulty arises from the need to make discrete selection decisions while accounting for interactions among borrowers~\cite{andersson2001, frey2001}. 
Even under simplified formulations, the number of feasible portfolios grows combinatorially as $C(N, K)=\binom{N}{K}$, making exhaustive search impractical for realistic problem sizes~\cite{garey1979, wolsey1998, conforti2014}. 
In addition, portfolio risk measures introduce pairwise interactions through covariance or correlation structures, further increasing computational complexity. 
A variety of solution approaches have therefore been investigated, including mixed-integer programming, stochastic optimization, and metaheuristic algorithms~\cite{bienstock1996, chang2000, Kolm2014, Mansini2015, crama2003, streichert2004}. 
Despite substantial progress, solving large-scale, constraint-rich portfolio optimization problems remains computationally demanding~\cite{Kolm2014, Mansini2015, okawara2023}.

Recent advances in quantum computing have motivated investigations of quantum algorithms for optimization problems in finance, including equity portfolio optimization, derivative pricing, risk analysis, and Monte Carlo simulation~\cite{orus2019, egger2020, herman2022, matsakos2024, Abbas2024, Yarkoni2022}. 
Particular attention has been devoted to hybrid quantum-classical algorithms designed for noisy intermediate-scale quantum (NISQ) devices, including quantum annealing, the Quantum Approximate Optimization Algorithm (QAOA), and Variational Quantum Algorithms (VQAs)~\cite{preskill2018, farhi2014, Albash2018, cerezo2021, bharti2022, Yarkoni2022}. 
These approaches typically reformulate optimization problems as Ising or Quadratic Unconstrained Binary Optimization (QUBO) models suitable for execution on quantum hardware~\cite{Lucas2014}. 
However, their applicability is often limited by the number of available qubits, hardware noise, and the measurement overhead required to evaluate objective functions on current devices~\cite{preskill2018, bharti2022, cerezo2021}. 
Moreover, many formulations employ a direct variable-to-qubit mapping, restricting the size of optimization problems that can be represented on existing hardware.

To address these limitations, qubit-efficient encoding strategies have recently been proposed that exploit the exponentially growing operator space of quantum systems to represent large optimization problems using comparatively few qubits. 
Among these approaches, Pauli Correlation Encoding (PCE) has emerged as a promising framework for large-scale combinatorial optimization~\cite{sciorilli2025, soloviev2026, PadinMartinez2026}. 
By encoding optimization variables in the expectation values of multi-qubit Pauli operators rather than assigning one qubit per variable, PCE enables problem representations whose variable count can substantially exceed the number of physical qubits. 
The correlations arising from PCE must be distinguished from the borrower-default correlations incorporated into the LPO formulation. 
The former are encoding-induced Pauli correlations among qubits, whereas the latter quantify statistical dependencies between borrower losses.

In this work, we investigate the application of PCE-based variational quantum optimization to large-scale LPO. 
Using a real-world microfinance dataset comprising 2012 borrowers and 26 months of historical loan data, we construct a correlation-aware portfolio optimization model based on borrower-level quantities, including exposure at default (EAD), loss given default (LGD), and probability of default (PD). 
The resulting formulation is expressed as a Quadratic Constrained Binary Optimization (QCBO) problem and subsequently transformed into a QUBO model suitable for VQAs. 
We evaluate the proposed framework through classical circuit simulations and experiments on IBM superconducting quantum processors, and benchmark the resulting solutions against Google OR-Tools~\cite{Google_ORTools} under comparable computational budgets.

The main contributions of this work are summarized as follows:
\begin{itemize}
\item We formulate a correlation-aware LPO model that jointly captures expected losses, loss variability, and borrower correlations while satisfying practical portfolio-selection constraints.
\item We develop a qubit-efficient quantum optimization framework based on PCE and VQAs, enabling LPO instances substantially larger than those feasible with conventional variable-to-qubit mappings.
\item We apply the proposed framework to a real-world microfinance dataset and investigate portfolio-optimization problems involving up to 1500 selection variables.
\item We benchmark the proposed framework against Google OR-Tools and evaluate its performance on IBM superconducting quantum hardware, providing an empirical assessment of PCE-based variational quantum optimization for large-scale LPO.
\end{itemize}

%=========================================================
\section{LPO Model}
\label{sec:lpo_model}

The objective of LPO is to select a subset of borrowers from a larger candidate pool such that portfolio risk is minimized while satisfying practical portfolio-selection constraints. Unlike traditional credit-scoring approaches, which evaluate borrowers individually, portfolio optimization must also account for dependencies among borrower losses. We therefore construct a correlation-aware optimization model that combines individual expected losses with portfolio-level risk arising from correlated borrower behavior.

To construct the LPO model, we assume that historical credit information is available for a set of $N$ borrowers over $T$ observation periods. In this work, these quantities are obtained from a real-world microfinance dataset comprising historical lending records. For each borrower, the available information includes Exposure at Default (EAD), Loss Given Default (LGD), and Credit Score (CS). These quantities are commonly used in credit-risk management and provide the fundamental inputs for estimating both borrower-level losses and portfolio-level risk~\cite{JPMcreditmetrics1997, creditriskplus1997, bluhm2010}.

We introduce the following notation and indices.
\begin{itemize}
	\item $i,j = 0,1,\cdots, N-1$: indices for borrowers, where $N$ is the number of borrowers.
	\item $t = 0,1,\cdots, T-1$: time-step index, where $T$ is the number of time periods.
\end{itemize}
Using the indexing scheme defined above, the borrower-level quantities are described as follows.
\begin{itemize}
	\item $\text{EAD}_{it}$: Exposure at Default (monetary value) of borrower $i$ at time $t$, obtained from product-level data. It is typically defined as the sum of the outstanding principal and accrued interest at a given time. In this work, we assume accrued interest is zero; therefore, EAD equals the outstanding principal only.
	
	\item $\text{LGD}_{it} = 1 - \text{Recovery Rate}_{it}$: Loss Given Default of borrower $i$ at time $t$, representing the fraction of exposure lost in the event of default. In this work, we assume
	\[
	\text{Recovery Rate}_{it} = 0 \quad \forall\, i,t,
	\]
	since we focus on unsecured micro-loans that typically do not have collateral. Therefore,
	\begin{equation}
		\label{LGD}
		\text{LGD}_{it} = 1 \quad \forall\, i,t.
	\end{equation}
	
	\item $\text{CS}_{it}$: Credit Score of borrower $i$ at time $t$, obtained from a credit bureau.
\end{itemize}
While EAD and LGD are obtained directly from historical loan records, the probability of default must be estimated from borrower characteristics. In many lending applications, credit scores are mapped to default probabilities through scorecard-based logistic regression models.
\begin{itemize}
	\item $\text{PD}_{it}$: Probability of Default of borrower $i$ at time $t$, obtained from an ML-based risk model. A simple and widely used approach estimates PD from the credit score using logistic regression \cite{lessmann2015} between the log-odds $\log\!\left(\frac{\text{PD}}{1-\text{PD}}\right)$ and the credit score $\text{CS}$:

	\begin{equation}
		\label{p}
		\text{PD}_{it} = \frac{1}{1 + e^{-(\beta_0 + \beta_1 \text{CS}_{it})}}.
	\end{equation}
	
	Here, $\beta_0$ is the intercept term representing the baseline portfolio risk, and $\beta_1$ is the slope coefficient controlling the sensitivity of PD to the credit score. In practice, $\beta_1$ is typically negative, implying that borrowers with higher credit scores have lower default probabilities.
	
	A commonly used industry setup assumes:
	\begin{itemize}
		\item Target Score = 600;
		\item Target Odds = 50:1 (50 good borrowers for every 1 default), corresponding to $\text{PD}=\tfrac{1}{51}$;
		\item PDO (Points to Double the Odds) = 20.
	\end{itemize}
	
	Under this setup, the logistic regression parameters are chosen as
	\begin{align}
		\label{betas}
		\beta_1 &\approx -\frac{\ln(2)}{\text{PDO}} \approx -0.0346, \\ \nonumber
		\beta_0 &\approx \ln\left(\frac{1}{50}\right) + \frac{\ln(2)}{\text{PDO}}\times 600 \approx 16.88,
	\end{align}
	which shifts the curve such that a credit score of 600 approximately corresponds to the target odds benchmark.
	
\end{itemize}

\subsection{Expected Loss (EL)}

Using the borrower-level quantities introduced above, we first construct a loss time series for each borrower over the observation horizon. These loss observations provide the foundation for estimating portfolio risk. We first define the expected loss, which represents a borrower’s average loss contribution and forms the first component of the portfolio objective.

\begin{figure}[!t]
    \centering
    \includegraphics[width=\linewidth]{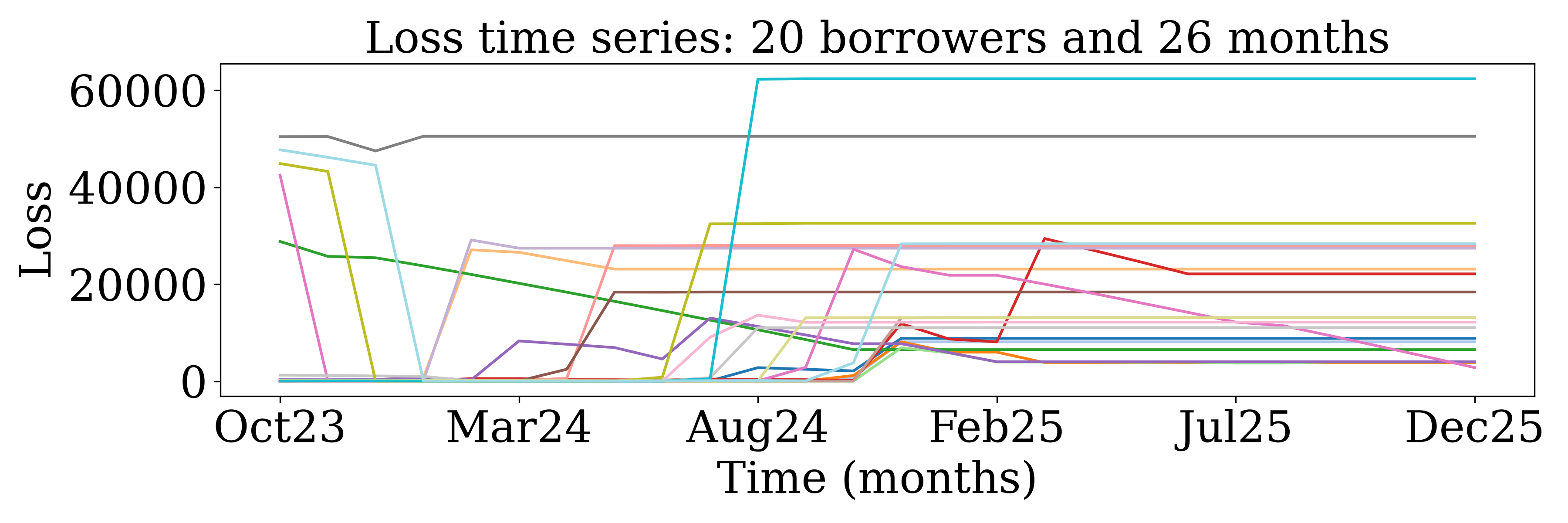}
    \caption{Loss time series ($L_{it}$) for 20 randomly selected borrowers in different colors. Each series contains 26 monthly observations from October 2023 to December 2025, excluding November 2024. For borrower $i$ at time $t$, the loss is proportional to the product of PD and EAD, where EAD is taken as the outstanding principal. As loans amortize, the outstanding principal decreases, reducing loss; however, rising PD can offset this effect. The observed loss trajectories reflect the combined influence of these two factors over time.}
    \label{fig:Lit}
\end{figure}

Using EAD, LGD, and PD, the loss associated with borrower $i$ at time $t$ is defined as 
\begin{equation} 
L_{it}=\text{EAD}_{it}\times \text{LGD}_{it}\times \text{PD}_{it}. 
\end{equation}
The collection of losses $L_{it}$ forms the borrower loss time series. 
Figure~\ref{fig:Lit} illustrates the loss time series $L_{it}$ for a representative sample of 20 borrowers. Each color represents a different borrower.

The expected loss for borrower $i$ is then~\cite{JPMcreditmetrics1997, creditriskplus1997, bluhm2010}
\begin{align}
	\label{EL_i}
	\mu_i &= \frac{1}{T}\sum_{t=0}^{T-1} L_{it}.
\end{align}
Using this, the expected loss of the portfolio, defined by the binary decision vector
$\boldsymbol{x} = (x_0, x_1, \cdots, x_{N-1})$, is
\begin{equation}
	\label{EL_PF}
	\text{EL}_\textsc{pf}(\boldsymbol{x})
	= \sum_{i=0}^{N-1} \mu_i x_i
	= \boldsymbol{\mu}^{\top}\boldsymbol{x},
\end{equation}
where
\begin{equation}
	\label{x}
x_i =
\begin{cases}
	1, & \text{if borrower } i \text{ is selected in the portfolio}, \\
	0, & \text{otherwise}.
\end{cases}
\end{equation}

\subsection{Unexpected Loss (UL)}

\begin{figure}[!t]
    \centering
    \includegraphics[width=0.95\columnwidth]{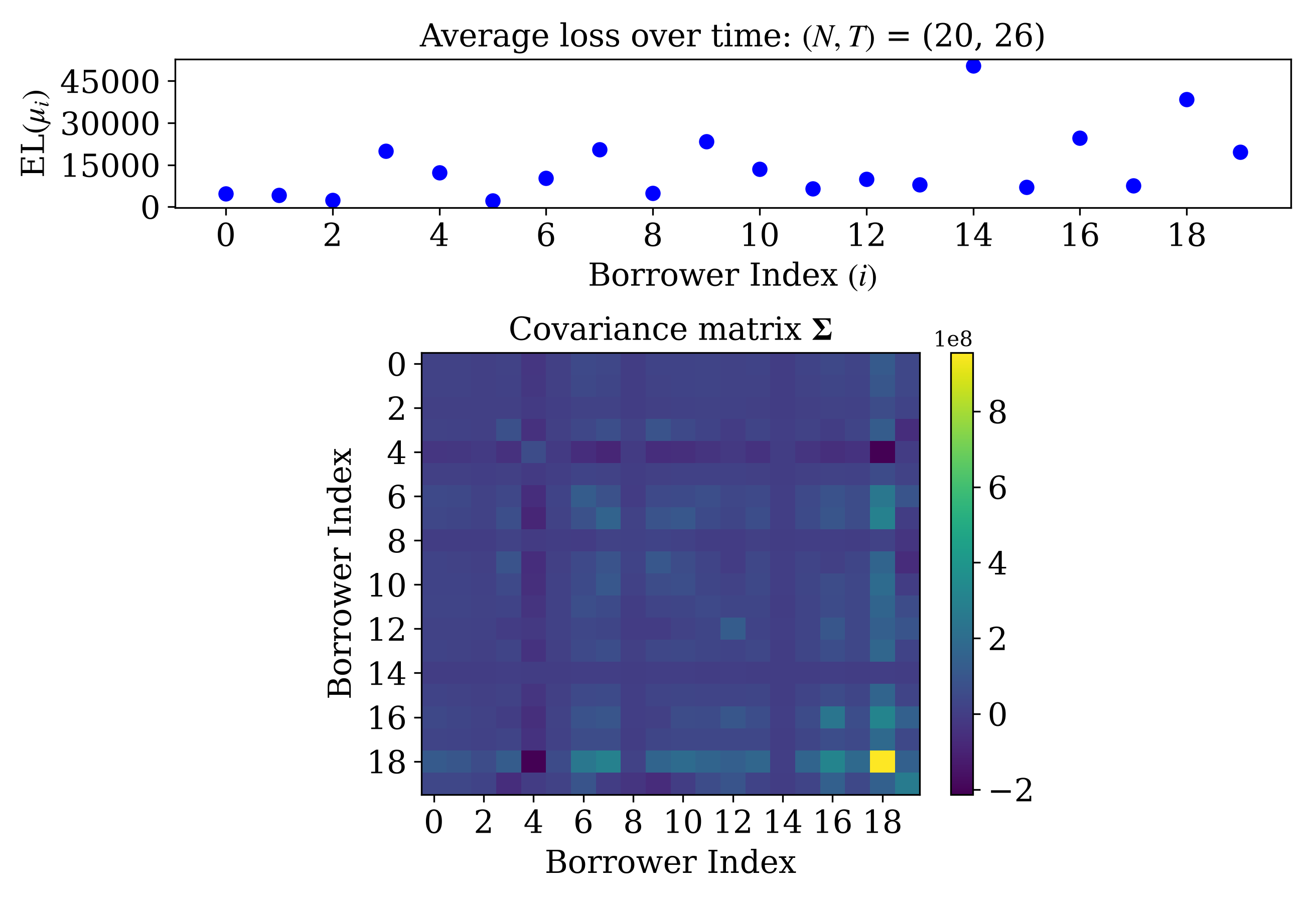}
    \caption{For the same set of 20 borrowers as Fig.~\ref{fig:Lit}, the top plot shows the expected losses $\boldsymbol{\mu}$, while the bottom heatmap presents the covariance matrix $\boldsymbol{\Sigma}$ of borrower losses. The expected losses capture the portfolio's expected risk, whereas the covariance matrix reflects the unexpected risk due to loss correlations across borrowers. Since the expected losses are point estimators, uncertainty intervals can be obtained from the diagonal entries of the covariance matrix; the square root of each diagonal element gives the corresponding standard deviation.}
    \label{fig:mu_sigma}
\end{figure}

While expected loss captures the average contribution of each borrower to portfolio risk, it does not account for the variability of losses or dependencies between borrowers. 
In practice, borrowers may be exposed to common economic, geographic, sectoral, or behavioral factors that can lead to correlated defaults. 
As a result, portfolios containing borrowers with similar risk characteristics may experience significantly larger losses during periods of financial stress. 
To capture this portfolio-level risk, we introduce a covariance-based measure of unexpected loss~\cite{JPMcreditmetrics1997, creditriskplus1997, bluhm2010}
\begin{equation}
	\label{UL^2}
	\text{UL}_\textsc{pf}^2(\boldsymbol{x})
	= \sum_{i=0}^{N-1}\sum_{j=0}^{N-1}
	x_i \,\Sigma_{ij} \,x_j
	= \boldsymbol{x}^{\top}\boldsymbol{\Sigma}\,\boldsymbol{x},
\end{equation}
where
\begin{equation}
	\Sigma_{ij}
	= \frac{1}{T-1}\sum_{t=0}^{T-1}
	(L_{it}-\mu_i)(L_{jt}-\mu_j)
\end{equation}
represents the covariance between losses of borrowers $i$ and $j$.
Figure~\ref{fig:mu_sigma} illustrates both the borrowers' expected losses $\boldsymbol{\mu}$ and the covariance matrix $\boldsymbol{\Sigma}$ for the representative subset of borrowers considered in Fig.~\ref{fig:Lit}. 
While the expected losses characterize the average contribution of individual borrowers to portfolio risk, the covariance matrix captures both the variability of individual losses and the interactions among borrowers arising from correlated loss behavior.

A practical challenge arises when the expected-loss and covariance terms are combined within a single optimization objective. 
As shown in Fig.~\ref{fig:mu_sigma}, the entries of $\boldsymbol{\mu}$ are typically of order $10^4$, whereas the entries of $\boldsymbol{\Sigma}$ are of order $10^8$. Furthermore, $\boldsymbol{\mu}$ has units of money, whereas $\boldsymbol{\Sigma}$ has units of money squared. Consequently, the two quantities are not directly comparable.

A natural way to express the unexpected loss in the same units as $\boldsymbol{\mu}$ would be to use the portfolio standard deviation, $\sqrt{\boldsymbol{x}^{\top}\boldsymbol{\Sigma}\,\boldsymbol{x}}$, derived from Eq.~\eqref{UL^2}. However, introducing the square root destroys the quadratic structure of the objective, making it unsuitable for the quadratic optimization formulations employed by both classical QUBO solvers and VQAs. Therefore, we define a modified covariance matrix, $\boldsymbol{\Tilde{\Sigma}}$, that preserves the quadratic form while reducing the scale mismatch and expressing the covariance terms in units of money.

Since $\boldsymbol{\Sigma}$ contains both positive and negative entries, an element-wise square root is not directly defined. We therefore take the square root of the absolute value of each entry while preserving its sign. The modified unexpected loss and covariance matrix are defined as
\begin{align}
\label{UL_modified}
\widetilde{\mathrm{UL}}_{\textsc{pf}}(\boldsymbol{x})
&=
\sum_{i=0}^{N-1}\sum_{j=0}^{N-1}
x_i \,
\Tilde{\Sigma}_{ij}
\, x_j,
\\[4pt]\nonumber
\Tilde{\Sigma}_{ij}
&=
\operatorname{sign}(\Sigma_{ij})
\sqrt{|\Sigma_{ij}|}.
\end{align}
This transformation preserves the sign of borrower loss correlations and reduces the scale mismatch between the expected-loss and covariance terms. Since $\boldsymbol{\mu}$ and $\boldsymbol{\Tilde{\Sigma}}$ are both expressed in units of money, they can be combined in a single objective. While $\boldsymbol{\Tilde{\Sigma}}$ is not a standard risk measure, it maintains a quadratic objective with comparable numerical scales.

The overall workflow may therefore be summarized as follows. $\mathrm{EAD}$, $\mathrm{LGD}$, and $\mathrm{CS}$ (used to derive $\mathrm{PD}$) constitute the raw input data for the LPO problem. 
These quantities are used to construct the borrower loss time series $L_{it}$. 
From the loss time series, we compute the expected loss $\boldsymbol{\mu}$ and the loss covariance matrix $\boldsymbol{\Sigma}$. 
These quantities then serve as inputs to the optimization model.

\subsection{Quadratic Constrained Binary Optimization (QCBO) Model}

The expected-loss vector $\boldsymbol{\mu}$ and covariance matrix $\boldsymbol{\Sigma}$ provide complementary descriptions of portfolio risk. 
Together, these quantities form the basis of the optimization problem described below.
The LPO problem aims to select $K$ borrowers out of $N$ by minimizing the following weighted combination of expected and modified unexpected losses, which we call the modified total loss:
\begin{equation}
	\label{TL_modified}
	\min_{\boldsymbol{x}} \;
	\underbrace{\alpha\,\mathrm{EL}_{\textsc{pf}}(\boldsymbol{x})
		+ \beta\,\widetilde{\mathrm{UL}}_{\textsc{pf}}(\boldsymbol{x})}_{\widetilde{\mathrm{TL}}_{\textsc{pf}}(\boldsymbol{x})},
\end{equation}
subject to the cardinality constraint
\begin{equation}
\label{cardinality_const}
	\sum_{i=0}^{N-1} x_i = K.
\end{equation}
Here, $\alpha \geq 0$ and $\beta \geq 0$ are weight parameters controlling the trade-off between expected and unexpected loss, respectively. Throughout this work, we set $\alpha=\beta=1$, assigning equal weight to both terms.
Without the proposed covariance modification, the total loss is defined as
\begin{equation}
\label{TL}
\mathrm{TL}_{\textsc{pf}}(\boldsymbol{x})
=
\alpha\,\mathrm{EL}_{\textsc{pf}}(\boldsymbol{x})
+
\beta\,\mathrm{UL}_{\textsc{pf}}(\boldsymbol{x}).
\end{equation}

\subsection{Quadratic Unconstrained Binary Optimization (QUBO) Model}

To solve the QCBO problem, defined by~\eqref{TL_modified} and~\eqref{cardinality_const} on a quantum computer, we first transform it into a QUBO problem of the form
\begin{equation}
\label{qubo}
	\min_{\boldsymbol{x}} \quad
	\boldsymbol{x}^{\top} Q\, \boldsymbol{x},
\end{equation}
where the $N\times N$ QUBO matrix
\begin{equation}
\label{Q_matrix}
	Q
	=
	\beta\,\boldsymbol{\Tilde{\Sigma}}
	+ \delta\,\mathbf{1}\mathbf{1}^{\top}
	+ \mathrm{diag}(\alpha\,\boldsymbol{\mu} - 2\delta K\,\mathbf{1})
\end{equation}
is obtained by solving
\begin{equation*}
	\alpha\,\text{EL}_\textsc{pf}(\boldsymbol{x})
	+ \beta\,\widetilde{\mathrm{UL}}_\textsc{pf}(\boldsymbol{x})
	+ \delta\left(\mathbf{1}^{\top}\boldsymbol{x} - K\right)^2
	=
	\boldsymbol{x}^{\top}Q\,\boldsymbol{x}
	+ \delta K^2,
\end{equation*}
thereby establishing the relationship between the QCBO and QUBO formulations. Here, $\delta \geq 0$ is the penalty parameter used to enforce the cardinality constraint, $\mathbf{1}$ denotes the $N$-dimensional column vector of ones, and $\mathrm{diag}(\cdot)$ represents the diagonal matrix formed from its vector argument.

In this work, we set $\delta=0$ and do not enforce the cardinality constraint through the QUBO penalty. Instead, the VQA learns a continuous score for each borrower, and the cardinality constraint is imposed exactly through a top-$K$ post-processing step described below the relation~\eqref{binarization}. Consequently, the quantum stage solves an unconstrained relaxed scoring problem, while feasibility is enforced deterministically through post-processing. More generally, the penalty term provides a flexible mechanism for incorporating portfolio selection constraints directly into the optimization formulation when required.

%=========================================================
\section{Classical Approaches}
\label{sec:Class_app}

\begin{figure}[!t]
    \centering
    \includegraphics[width=0.8\linewidth]{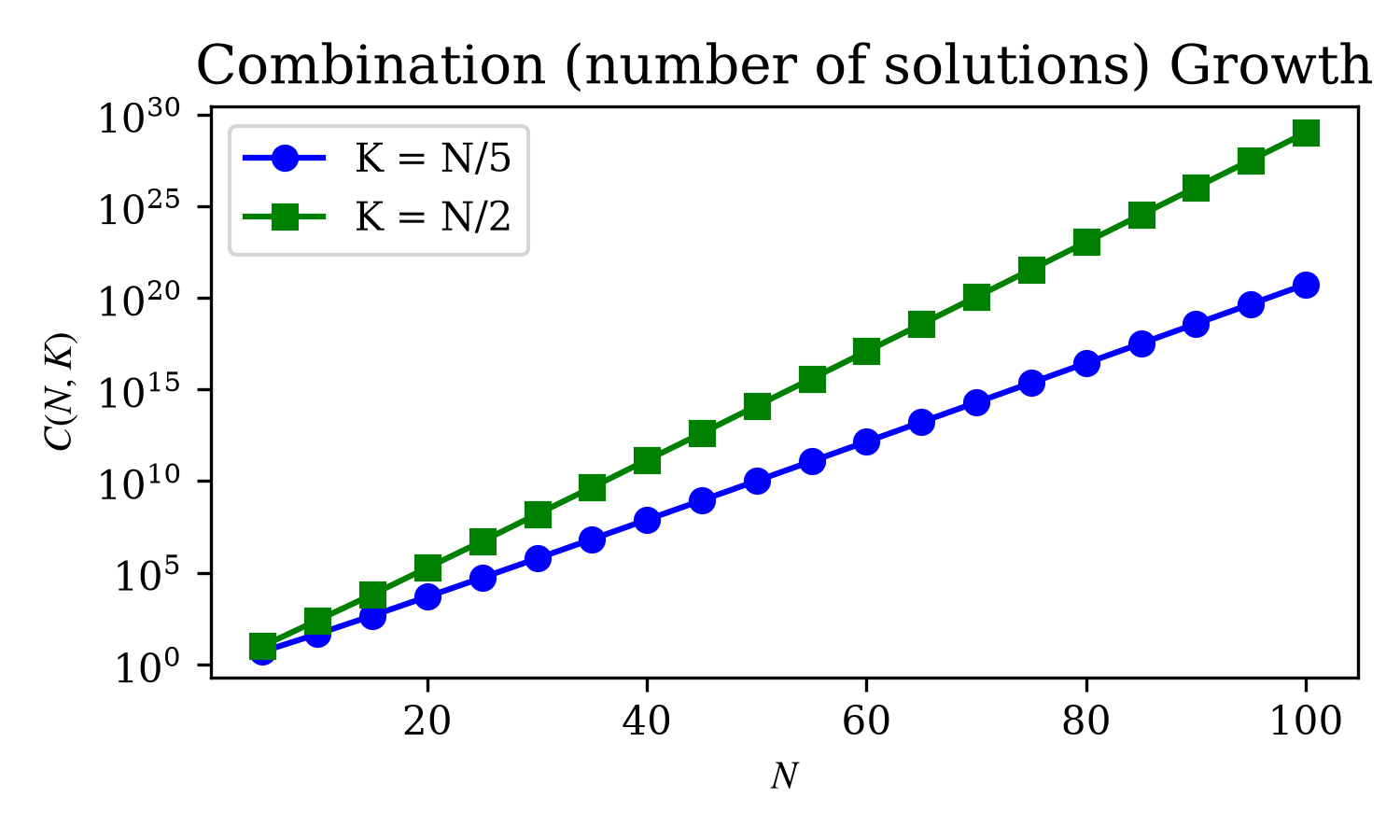}
    \caption{Growth of the LPO solution space with the number of borrowers, $N$, for $K=N/2$ (50\% selection) and $K=N/5$ (20\% selection).}
    \label{fig:CNK}
\end{figure}

A feasible solution to the LPO problem defined by~\eqref{TL_modified} and~\eqref{cardinality_const} is an $N$-dimensional binary vector $\boldsymbol{x}$ containing exactly $K$ ones and $N-K$ zeros. The number of feasible solutions is given by the binomial coefficient
\begin{equation}
C(N,K) = \binom{N}{K}= \frac{N!}{K!(N-K)!},
\end{equation}
which grows combinatorially with $N$. Consequently, the size of the solution space increases rapidly even for moderate values of $N$ and $K$, as illustrated in Fig.~\ref{fig:CNK}.

For small problem instances, the optimal solution can be obtained through brute-force search by enumerating all feasible binary vectors, evaluating the total loss $\mathrm{TL}_{\textsc{pf}}(\boldsymbol{x})$ for each portfolio, and selecting the portfolio with the minimum loss value. However, the combinatorial growth of the solution space renders this approach computationally intractable as $N$ increases.

\subsection{Linear Constrained Binary Optimization (LCBO) Model}

For moderate problem sizes, exact optimization solvers such as Google OR-Tools~\cite{Google_ORTools} can be used. However, the objective function in~\eqref{TL_modified} is quadratic, whereas the CP-SAT solver requires a linear objective. To address this, we introduce auxiliary binary variables $z_{ij}$ to linearize the quadratic terms in Eq.~\eqref{UL_modified}, resulting in the following
\begin{equation}
\label{ortool_obj}
\min_{\boldsymbol{x}, \boldsymbol{z}}
\quad
\alpha\,\mathrm{EL}_{\textsc{pf}}(\boldsymbol{x})
+
\beta\,
\sum_{i=0}^{N-1}\sum_{j=0}^{N-1}
\Tilde{\Sigma}_{ij}\, z_{ij},
\end{equation}
subject to
\begin{align}
\label{ortool_consts}
\sum_{i=0}^{N-1} x_i &= K, \nonumber\\
z_{ij} &\le x_i \qquad \forall\, i,j,\\
z_{ij} &\le x_j \qquad \forall\, i,j, \nonumber\\
x_i + x_j - 1 &\le z_{ij} \qquad \forall\, i,j. \nonumber
\end{align}
Here, the last three constraints ensure that $z_{ij}=x_i x_j$. The vector $\boldsymbol{z}=(\cdots,z_{ij},\cdots)$ contains $N^2$ binary variables.

As the brute-force algorithm explicitly evaluates every feasible portfolio, its time complexity is proportional to the number of feasible solutions, $C(N,K)$.
For example, when $K=N/2$,
\[
C(N,N/2) \approx \frac{2^N}{\sqrt{\pi N/2}},
\]
using Stirling's approximation. In contrast, OR-Tools CP-SAT employs branch-and-bound, constraint propagation, and SAT-based pruning to avoid explicit enumeration of all feasible solutions. Although the underlying optimization problem remains NP-hard and exhibits exponential worst-case complexity, CP-SAT can solve moderate-sized instances much more efficiently in practice.

Through Pauli Correlation Encoding (PCE), quantum computing offers the possibility of representing and optimizing certain large-scale problems using substantially fewer qubits than binary decision variables~\cite{sciorilli2025, soloviev2026, PadinMartinez2026}. In the following sections, we employ a variant of this encoding within a VQA framework to solve the LPO problem.

%=========================================================
\section{Variational Quantum Algorithm}
\label{sec:VQA}

To solve the QUBO formulation of the LPO problem introduced in Sec.~\ref{sec:lpo_model}, we employ a Variational Quantum Algorithm (VQA). VQAs combine quantum state preparation and measurement with classical optimization, making them particularly suitable for NISQ devices~\cite{cerezo2021, bharti2022}. In the proposed framework, the VQA optimizes a relaxed representation of the portfolio-selection problem, while a classical post-processing step enforces feasibility with respect to the portfolio cardinality constraint.
The quantum component consists of two main parts: (i) a parameterized quantum circuit (PQC) for preparing a variational quantum state, and (ii) quantum measurements for estimating expectation values required to evaluate the relaxed QUBO objective function given in~\eqref{relaxed_QUBO}. A classical optimizer then uses these expectation values to iteratively update the circuit parameters and minimize the relaxed objective function.

\subsection{Parameterized Quantum Circuit}

\begin{figure}[!t]
    \centering
    \includegraphics[width=\linewidth]{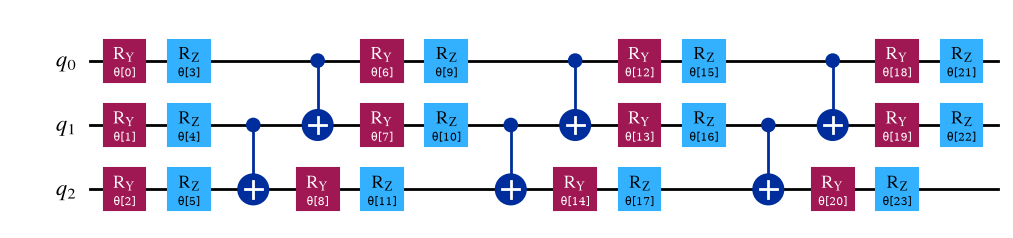}
    \caption{The PQC corresponding to the \texttt{EfficientSU2} ansatz available in Qiskit~\cite{qiskit2024, Kandala2017} and defined in Eq.\eqref{theta_ket}. Here, the number of qubits is $n=3$, and the number of layers is $\mathscr{L}=3$. Each layer consists of parameterized single-qubit rotation gates $R_y$ and $R_z$, represented by red and blue rectangular blocks, respectively, followed by entangling controlled-NOT ($\mathrm{CX}$) gates acting between qubits $a$ and $a+1$, represented by a control dot connected to a target $\oplus$ symbol. The qubits, labeled $q_0,\cdots,q_a,\cdots,q_{n-1}$, are represented by horizontal wires. Here, each variational parameter is denoted by $\theta$ with a single index, rather than the two-index notation used in Eq.~\eqref{theta}.}
    \label{fig:EffSU2_PQC}
\end{figure}

We use the hardware-efficient \texttt{EfficientSU2} ansatz~\cite{qiskit2024, Kandala2017} available in Qiskit as the PQC (see Fig.~\ref{fig:EffSU2_PQC}). The circuit consists of alternating layers of parameterized single-qubit rotation gates and entangling gates. The single-qubit rotations are generated from the SU(2) group using Pauli rotation gates such as $R_y$ and $R_z$, while entanglement between control qubit $a$ and target qubit $b$ is introduced using controlled-NOT $\mathrm{CX}(a,b)$ gates. This layered structure enables the circuit to represent complex quantum states while remaining compatible with near-term quantum hardware.
For an $n$-qubit system, the PQC prepares a variational quantum state
\begin{align}
	\label{theta_ket}
	|\boldsymbol{\theta}\rangle
	&=
	U(\boldsymbol{\theta})|0\rangle^{\otimes n}, \\ \nonumber
	U(\boldsymbol{\theta})
	&=
    \bigotimes_{a=0}^{n-1}
    R_z(\vartheta_a^{\mathscr{L}}) 
	R_y(\theta_a^{\mathscr{L}})\\ \nonumber
    &\quad
	\prod_{l=0}^{\mathscr{L}-1}
	\left[
	\left(
	\bigotimes_{a=0}^{n-2}
	\mathrm{CX}(a,a+1)
	\right)
	\left(
	\bigotimes_{a=0}^{n-1}
    R_z(\vartheta_a^{l})
	R_y(\theta_a^{l})
	\right)
	\right],
\end{align}
where the unitary operator $U(\boldsymbol{\theta})$ denotes the PQC with trainable parameters
\begin{equation}
	\label{theta}
	\boldsymbol{\theta} := (\cdots,\theta_a^{l},\vartheta_a^{l},\cdots),
\end{equation}
$\mathscr{L}$ is the number of circuit layers, and the subscript $a$ and superscript $l$ denote the qubit index and layer index, respectively.

\subsection{Pauli Operator Measurement}
The objective function is evaluated using expectation values of Pauli operators. Ignoring the phase factors, the single-qubit Pauli group 
\begin{equation}
	\label{IXYZ}
	\mathcal{P}:=\{I, X, Y, Z\}
\end{equation}
consists of the identity operator $I$ and the Pauli operators $X$, $Y$, and $Z$.
For an $n$-qubit system, the tensor products of operators from $\mathcal{P}$ form an orthogonal operator basis of size $4^n$ under the Hilbert-Schmidt inner product. Excluding the identity operator $I^{\otimes n}$, there are
$4^n - 1$ non-trivial $n$-qubit Pauli operators. Accordingly, representing $N$ binary variables requires the minimum number of qubits satisfying
$N \leq 4^n - 1,$
which gives
\begin{equation}
\label{n}
	n = \left\lceil \log_4(N+1) \right\rceil + m,
\end{equation}
where the additional parameter $m\geq 0$ (set to $0$ throughout) accounts for extra qubits that can be added to increase the expressibility of the variational quantum state.

We then randomly select $N$ non-identity Pauli operators
\begin{equation} 
    \label{Pauli_ops_N} 
    \{P_0,P_1,\cdots,P_{N-1}\}, \quad P_i= \bigotimes_{a=0}^{n-1}\sigma_a^{(i)}, \quad \sigma_a^{(i)}\in\mathcal P, 
\end{equation} 
and associate them with the $N$ binary decision variables of the LPO problem. The variational quantum state $|\boldsymbol{\theta}\rangle$ is measured with respect to these operators to evaluate the optimization objective.

\subsection{Classical Optimization in the VQA}

A classical optimizer, such as the COBYLA optimizer provided in SciPy~\cite{cobyla-method, 2020SciPy-NMeth}, solves the following optimization problem:
\begin{align}
	\label{relaxed_QUBO}
	&\min_{\boldsymbol{\theta}} \quad
	\boldsymbol{x}(\boldsymbol{\theta})^{\top}
	Q\,
	\boldsymbol{x}(\boldsymbol{\theta}), \\ \nonumber
	\boldsymbol{x}(\boldsymbol{\theta})
	&=
	\left(
	x_0(\boldsymbol{\theta}),
	\cdots,
	x_{N-1}(\boldsymbol{\theta})
	\right), \\ \nonumber
	x_i(\boldsymbol{\theta}; \gamma)
	&=
	\frac{
		1 + \tanh\left(
		\gamma\;
		\langle\boldsymbol{\theta}|P_i|\boldsymbol{\theta}\rangle
		\right)
	}{2}
	\in [0,1],
\end{align}
where the Pauli expectation values
\begin{equation}
	\label{expt_pauli}
	\langle\boldsymbol{\theta}|P_i|\boldsymbol{\theta}\rangle \in [-1,1]
\end{equation}
are obtained from quantum measurements.
Since Pauli operators are generally non-commuting, their expectation values are not independent and satisfy polynomial equality and inequality constraints~\cite{Kimura2003, Byrd2003, Sehrawat2020}. For example, the expectation values of all the Pauli operators cannot simultaneously attain the value $1$. Because the expectation values are constrained and cannot generally attain independent binary limits, we have an annealing parameter $\gamma > 0$.
As $\gamma$ increases, the variables gradually approach binary values:
\begin{equation}
\label{binarization}
x_i(\boldsymbol{\theta}; \gamma)
\rightarrow
\begin{cases}
0 & \mbox{if}\ x_i(\boldsymbol{\theta}; \gamma) < \tfrac{1}{2}\\[2mm]
1 & \mbox{if}\ x_i(\boldsymbol{\theta}; \gamma) \geq \tfrac{1}{2}
\end{cases}
\end{equation}
thereby yielding a discrete solution.
The role of the parameter $\gamma$ has recently been investigated in the context of budget-constrained MinCut problems, where progressively increasing $\gamma$ was shown to improve binarization, constraint satisfaction, and overall solution quality \cite{PadinMartinez2026}.

The optimization is performed for a fixed number of epochs, over which $\gamma$ is increased geometrically from $0.3$ to $50$. 
To stabilize training, the best parameter vector (determined by the minimum QUBO cost evaluated on the discrete solution) is updated using an exponential moving average whenever the discrete-solution QUBO cost obtained from the current parameter vector is within $10\%$ of the best solution found so far. After all epochs are completed, the best parameter vector, denoted by $\boldsymbol{\theta}^{*}$, is used to generate the relaxed decision vector $\boldsymbol{x}^{*}:=\boldsymbol{x}(\boldsymbol{\theta}^{*})$.

However, binarizing $\boldsymbol{x}^{*}$ using a fixed threshold, such as $1/2$ as in~\eqref{binarization} does not generally guarantee satisfaction of the cardinality constraint Eq.~\eqref{cardinality_const} even for $\delta > 0$ in Eq.~\eqref{Q_matrix}. To address this issue, we post-process the obtained probabilities $\boldsymbol{x}^{*}$ by setting the top $K$ probabilities to $1$ and the remaining $N-K$ probabilities to $0$, yielding the discrete solution $\boldsymbol{x}_{\textsc{vqa}}$. This procedure automatically enforces the cardinality constraint while preserving the ranking information learned by the quantum model.

All optimization methods described above, except brute-force search, operate on the modified total loss objective $\widetilde{\mathrm{TL}}_{\textsc{pf}}(\boldsymbol{x})$ defined in~\eqref{TL_modified}. 
This modification is introduced to reduce the scale mismatch between the expected-loss and covariance terms while preserving a quadratic objective function. However, the quality of all solutions reported in the remainder of this paper is evaluated using the original total loss $\mathrm{TL}_{\textsc{pf}}(\boldsymbol{x})$ defined in Eq.~\eqref{TL}. 

%=========================================================
\section{LPO Results}
\label{sec:LPO_results}

\begin{figure}[!t]
    \centering
    \includegraphics[width=\columnwidth]{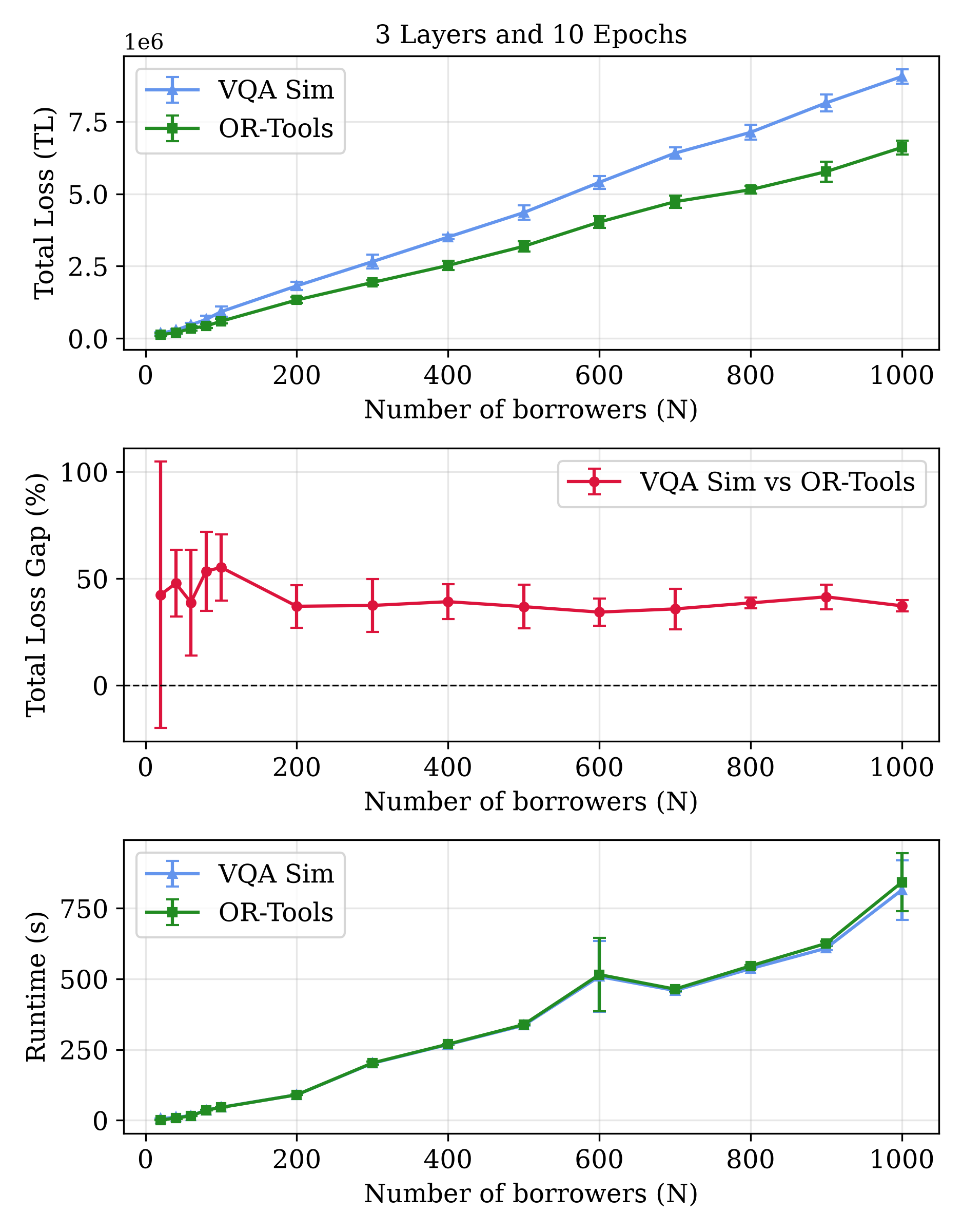}
    \caption{Scaling behavior for ${K=N/2}$ using the full PCE described in \eqref{Pauli_ops_N}. The top, middle, and bottom panels show the portfolio loss, relative gap, and runtime as functions of the problem size $N$. For each value of $N$, the marker denotes the mean over five random instances sampled from the dataset of ${2012}$ borrowers, while the error bars represent the corresponding $95\%$ confidence intervals. The top panel compares the portfolio total loss obtained by VQA simulation in Qiskit and OR-Tools applied to the LCBO model described in Sec.~\ref{sec:Class_app}. The middle panel shows the relative gap (\%) between the solution qualities of VQA and OR-Tools, as defined in Eq.~\eqref{gap}. The bottom panel shows the VQA runtime (in seconds) as a function of $N$. For each instance, the OR-Tools time limit is set equal to the corresponding VQA runtime.}
\label{fig:scaling_N20_1000}
\end{figure}

\subsection{Scalability Analysis}\label{sec:scalability}

To evaluate the proposed framework, we investigate two complementary questions. First, how does the solution quality achieved by the proposed VQA simulation compare with that obtained using Google OR-Tools~\cite{Google_ORTools} as the problem size increases? Second, how closely do the solutions obtained by executing the VQA on IBM quantum hardware agree with those produced by its classical simulations? Together, these experiments provide insight into both the scalability of the proposed framework and its practical performance on current quantum processors. Throughout this section, we set ${K=N/2}$.

Our dataset comprises $2012$ borrowers with $26$ months of historical loss data from real-world microfinance loans. 
For the scalability study, we consider problem sizes
\begin{equation}
\label{Ns}
N \in \{20,40,60,80,100,200,\cdots,900,1000\}.
\end{equation}
For each value of $N$, we randomly sample five problem instances from the dataset. Each instance is first solved using the VQA framework described in the previous section and implemented with the Qiskit simulator, and then solved using Google OR-Tools applied to the LCBO model defined by~\eqref{ortool_obj} and~\eqref{ortool_consts}.

\begin{figure}[!t]
    \centering
    \includegraphics[width=\columnwidth]{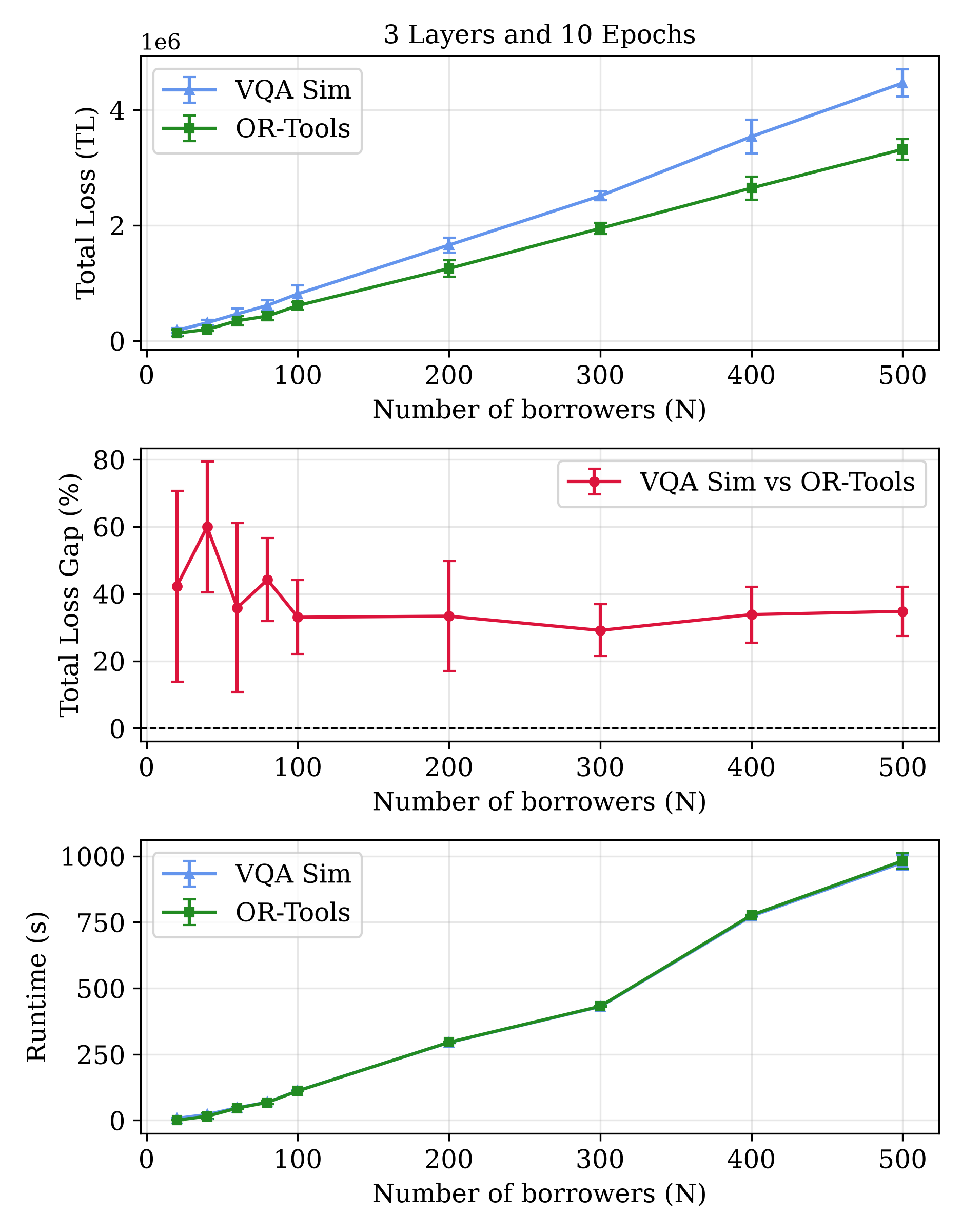}
    \caption{Scaling behavior for $K=N/2$ using the three-subgroup PCE described around \eqref{nprime}. The quantities shown are the same as those in Fig.~\ref{fig:scaling_N20_1000}, but for problem sizes $N=\{20,40,60,80,100,200,300,400,500\}$.}
\label{fig:scaling_N20_500_3PG}
\end{figure}

For the VQA simulation, we employ a PQC shown in Fig.~\ref{fig:EffSU2_PQC} with 3 layers and run the optimization for 10 epochs. 
For each instance $s$, the VQA produces a solution $\boldsymbol{x}^{\,s}_{\textsc{vqa}}$ with a runtime $t^{\,s}_{\textsc{vqa}}$. 
The same runtime is then used as the time limit for OR-Tools, which returns a solution $\boldsymbol{x}^{\,s}_{\textsc{ort}}$. 
Thus, for each instance, we obtain one solution from VQA and one solution from OR-Tools under the same computational budget. 
The corresponding portfolio losses using Eq.~\eqref{TL} and their relative gap are computed as
\begin{align}
\label{gap}
\mathrm{TL}^{\,s}_{\textsc{vqa}}
&=
\mathrm{TL}_{\textsc{pf}}
\!\left(
\boldsymbol{x}^{\,s}_{\textsc{vqa}}
\right),
\nonumber\\
\mathrm{TL}^{\,s}_{\textsc{ort}}
&=
\mathrm{TL}_{\textsc{pf}}
\!\left(
\boldsymbol{x}^{\,s}_{\textsc{ort}}
\right),
\\
\mathrm{Gap}^{\,s}
&=
\frac{
\mathrm{TL}^{\,s}_{\textsc{vqa}}
-
\mathrm{TL}^{\,s}_{\textsc{ort}}
}{
\mathrm{TL}^{\,s}_{\textsc{ort}}
}
\times 100.
\nonumber
\end{align}
For each problem size $N$, the results are averaged over five instances, and the corresponding $95\%$ confidence intervals are reported as error bars. 
Figure~\ref{fig:scaling_N20_1000} presents the first set of results, showing the scaling behavior of the portfolio loss, relative gap, and runtime.

The top panel compares the total portfolio loss obtained by VQA simulation and OR-Tools. As expected, the portfolio loss increases with $N$ since larger portfolios contain more selected borrowers. The slope of each curve reflects the quality of the corresponding solver: a lower slope indicates a lower loss and hence a better portfolio. The OR-Tools curve consistently lies below the VQA curve, indicating higher-quality solutions. The difference between the two solution qualities is quantified by the relative gap shown in the middle panel.

The middle panel shows the relative gap between the VQA and OR-Tools solution qualities. The gap remains close to $40\%$ across the entire range of problem sizes given in~\eqref{Ns}, indicating that the VQA consistently produces solutions whose objective values are approximately $40\%$ higher than those obtained by OR-Tools when both methods are allocated the same computational budget.

The bottom panel shows the VQA runtime as a function of $N$. With the number of PQC layers fixed at three and the number of optimization epochs fixed at ten, the average runtime increases from approximately $5$ seconds for $N=20$ to about ${800}$ seconds for $N=1000$. Since the OR-Tools time limit is set equal to the VQA runtime for each instance, the runtime plot reflects the computational cost of the VQA and does not represent the standalone runtime scaling of OR-Tools.

The execution of the \emph{full} PCE encoding, described in \eqref{Pauli_ops_N}, on quantum hardware can incur significant measurement overhead, as the selected Pauli operators may span as many as \(2^n+1\) commuting subgroups. To reduce the number of commuting subgroups, and hence the number of measurement settings required for hardware execution, we consider three mutually commuting sets of Pauli operators generated from tensor products of \(\{I,X\}\), \(\{I,Y\}\), and \(\{I,Z\}\), respectively. Excluding the identity operator, each set contains \(2^{n'}-1\) operators, where \(n'\) denotes the number of qubits. Consequently, this encoding provides a total of \(3(2^{n'}-1)\) non-trivial Pauli operators.

To represent an optimization problem with $N$ binary variables, we choose the minimum number of qubits satisfying
$N \leq 3(2^{n'}-1)$,
which yields
\begin{equation}
\label{nprime}
n'
=
\left\lceil
\log_2\!\left(\frac{N}{3}+1\right)
\right\rceil.
\end{equation}
Throughout this study, no additional qubits are introduced beyond this minimum requirement.

To assess the impact of this restriction on optimization performance, we repeat the scalability analysis of Fig.~\ref{fig:scaling_N20_1000} using the \emph{three-subgroup} PCE. Figure~\ref{fig:scaling_N20_500_3PG} presents the resulting portfolio losses, relative gaps, and runtimes. The observed trends closely mirror those obtained with the full PCE. Across the range of problem sizes considered, the relative gap with respect to Google OR-Tools remains approximately $40\%$, indicating that restricting the encoding to three commuting Pauli subgroups does not significantly degrade solution quality. Despite relying on a substantially smaller operator pool, the three-subgroup PCE achieves performance comparable to that of the full PCE while being considerably more suitable for quantum hardware execution.
However, the classical simulation runtimes are noticeably longer than those obtained with the full PCE, owing to the larger number of qubits required by the three-subgroup PCE, i.e., $n' \geq n$ for a given problem size $N$.

\subsection{Quantum Hardware Experiments}

\begin{figure}[!t] 
\centering 
\includegraphics[width=\linewidth]{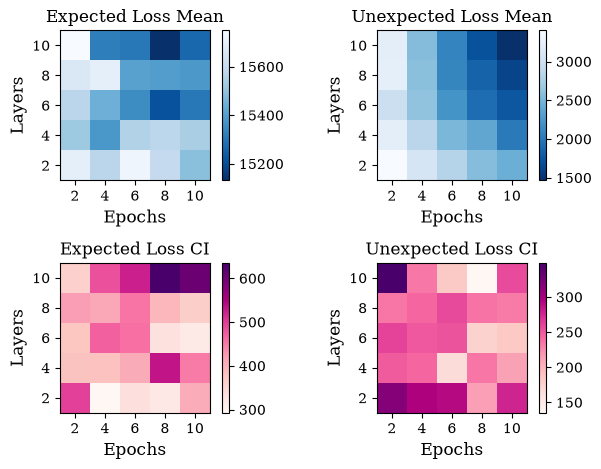} 
\caption{Sensitivity of the three-subgroup PCE to PQC depth and optimization epochs for $(N,K)=(500,250)$. Results are averaged over ten independently sampled LPO instances. The upper panels show the mean expected and unexpected portfolio losses, while the lower panels show the corresponding $95\%$ confidence-interval half-widths associated with the estimated means.} 
\label{fig:depth_epoch} 
\end{figure}

We now present the second set of results, focusing on the execution of the proposed framework on IBM quantum hardware. The experiments employ the three-subgroup PCE introduced in Sec.~\ref{sec:scalability}, which reduces measurement overhead by restricting the operator pool to three mutually commuting Pauli subgroups.

Before performing the hardware experiments, we investigate the sensitivity of the three-subgroup PCE to the PQC depth and the number of optimization epochs. Since the scalability study in Fig.~\ref{fig:scaling_N20_500_3PG} considers problem sizes ranging from $N=20$ to $N=500$, we select representative largest instances with $(N,K)=(500, 250)$ for this analysis. Ten independent LPO instances are sampled from the dataset and solved for each combination of circuit depth and optimization epochs. The corresponding expected and unexpected losses are then averaged across the ten instances.

Figure~\ref{fig:depth_epoch} summarizes the results with the corresponding $95\%$ confidence intervals. Increasing the number of layers and optimization epochs generally improves solution quality, with the lowest unexpected loss obtained using ten layers and ten epochs. However, hardware-execution costs also increase with both quantities. For the \texttt{EfficientSU2} ansatz, the total QPU usage time of our VQA scales approximately as
\begin{equation}
\label{QPU_time}
    \Bigl((2n'(\mathscr{L}+1)+3) \times \text{no. epochs}+1\Bigr)\,t_{\mathrm{QPU}},
\end{equation} 
where $n'$ is the number of qubits, $\mathscr{L}$ is the number of PQC layers, and $t_{\mathrm{QPU}}$ denotes the execution time required to estimate all the $N$ Pauli expectation values.

Consequently, greater circuit depth, more optimization iterations, and more commuting Pauli groups lead to substantially higher QPU costs. To balance solution quality and hardware resource requirements, the experiments reported below use ${10}$ PQC layers and ${6}$ optimization epochs.

\begin{figure*}[!t]
    \centering
    \includegraphics[width=\textwidth]{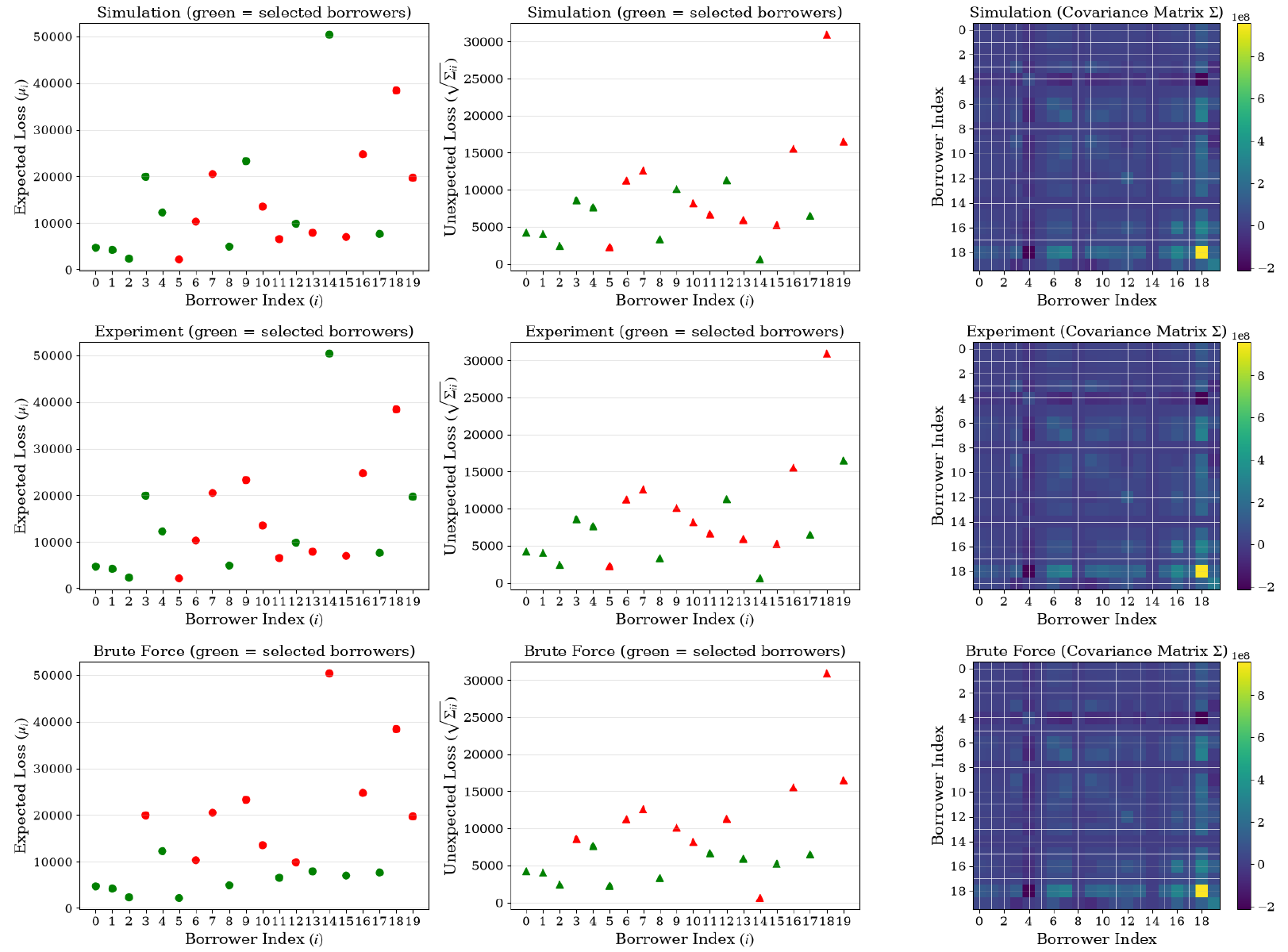}
    \caption{Comparison of portfolio solutions for a representative instance with $N=20$ and $K=10$. The top, middle, and bottom rows show the solutions obtained from VQA simulation, VQA execution on IBM quantum hardware, and brute-force search, respectively. Since brute-force search exhaustively evaluates all feasible portfolios, the bottom row represents the optimal solution. In each row, the left, middle, and right panels display the expected loss, unexpected loss (standard deviation), and covariance matrix of borrower losses, respectively. Selected borrowers are highlighted in green and excluded borrowers in red. A high-quality portfolio is characterized by borrowers with low expected loss, low unexpected loss, and low pairwise covariance. In the covariance matrix, horizontal and vertical white lines mark the selected borrowers; the corresponding submatrix lies predominantly in low-covariance (blue) regions, indicating effective diversification. Overall, the VQA simulation and hardware solutions resemble the optimal brute-force solution across all three risk perspectives.}
    \label{fig:N20_K10_VQA_BF}
\end{figure*}

We begin with a representative instance of size $(N, K)=(20,10)$, for which brute-force benchmarking remains feasible, and the optimal portfolio can therefore be determined exactly. Figure~\ref{fig:N20_K10_VQA_BF} compares the solutions obtained from VQA simulation, VQA execution on IBM quantum hardware, and brute-force search. The figure visualizes each solution from three complementary perspectives: individual expected losses, individual unexpected losses (standard deviations), and the covariance matrix of borrower losses. In the plots of expected and unexpected losses, the selected borrowers are concentrated among the lower-loss candidates, indicating favorable individual risk characteristics. In the covariance matrix, the selected borrowers are predominantly located in low-covariance regions, reflecting weak correlations and hence better portfolio diversification.

\begin{figure}[!t]
    \centering
    \includegraphics[width=\columnwidth]{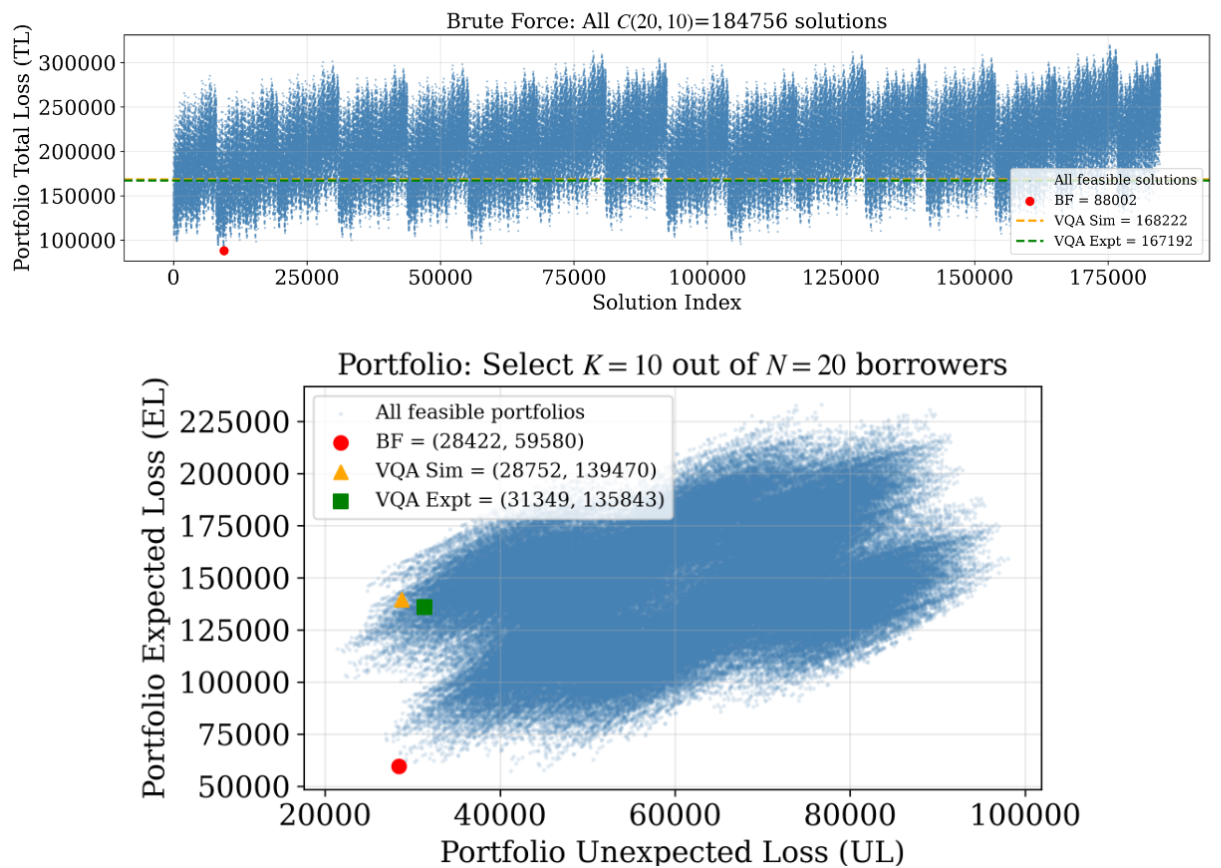}
    \caption{Portfolio-level comparison for a representative instance with $(N,K)=(20,10)$. The top panel shows the total loss of all $C(20,10)=184{,}756$ feasible portfolios represented by blue dots. The red marker denotes the optimal brute-force solution, while the orange and green markers correspond to VQA simulation and VQA execution on IBM quantum hardware, respectively. The bottom panel shows the same portfolios in the unexpected-loss–expected-loss plane, analogous to an efficient frontier, with unexpected loss on the horizontal axis and expected loss on the vertical axis. Portfolios near the lower-left corner exhibit both low expected and low unexpected loss and are therefore preferred.}
    \label{fig:N20_K10_VQA_BF_eff_front}
\end{figure}

Figure~\ref{fig:N20_K10_VQA_BF_eff_front} presents portfolio-level results for the same instance. The top panel shows the total losses of all feasible portfolios and highlights the solutions obtained from brute-force search, VQA simulation, and VQA execution on IBM quantum hardware. The VQA simulation and hardware solutions lie close to each other, although both deviate from the optimal brute-force solution.

The bottom panel provides an efficient-frontier-like view of the same portfolios. The horizontal and vertical axes represent the portfolio unexpected loss (standard deviation) and expected loss, respectively. All three solutions lie toward the left side of the plot, indicating relatively low unexpected loss. The primary difference arises in the expected loss, which determines the separation between the solutions. The brute-force solution lies closest to the lower-left corner, corresponding to the most favorable risk profile. The VQA simulation and hardware solutions show moderate deviations from this optimum.

The results are obtained with $\alpha=\beta=1$. Since the expected-loss term contains $O(K)$ active contributions, whereas the quadratic covariance term contains up to $O(K^2)$ pairwise contributions, their relative influence can vary with the portfolio size $K$, even after the covariance transformation introduced in Eq.~\eqref{UL_modified}. This motivates considering a balanced weighting, 
\begin{equation} 
\label{beta_balance}
\beta_{\mathrm{balance}} = 
\frac{\boldsymbol{\mu}^{\top}\boldsymbol{1}} {\boldsymbol{1}^{\top}\boldsymbol{\tilde{\Sigma}}\,\boldsymbol{1}}, \end{equation} 
which equalizes the aggregate scales of the expected-loss and covariance terms. Preliminary observations suggest that this choice may yield VQA solutions closer to the optimal brute-force solution. A systematic investigation is left for future work.

We now address the second question posed at the beginning of this section: how closely do the solutions obtained on IBM quantum hardware match those produced by classical VQA simulations on Qiskit?

We sample four LPO instances with problem sizes 
\begin{equation}
\label{Ns_expt}
N \in \{20,400,1000,1500\}\,,
\end{equation}
considering the case \(K=N/2\). Each instance was solved using both VQA simulation and VQA experiments on IBM quantum hardware. 
For a fair comparison, the same VQA architecture, consisting of 10 ansatz layers and 6 optimization epochs, was used in both cases. In addition, the same initial parameters, three-subgroup PCE, and all other settings described in Sec.~\ref{sec:VQA} were kept identical across the simulations and hardware experiments.

\begin{figure}[!t]
    \centering
    \includegraphics[width=\columnwidth]{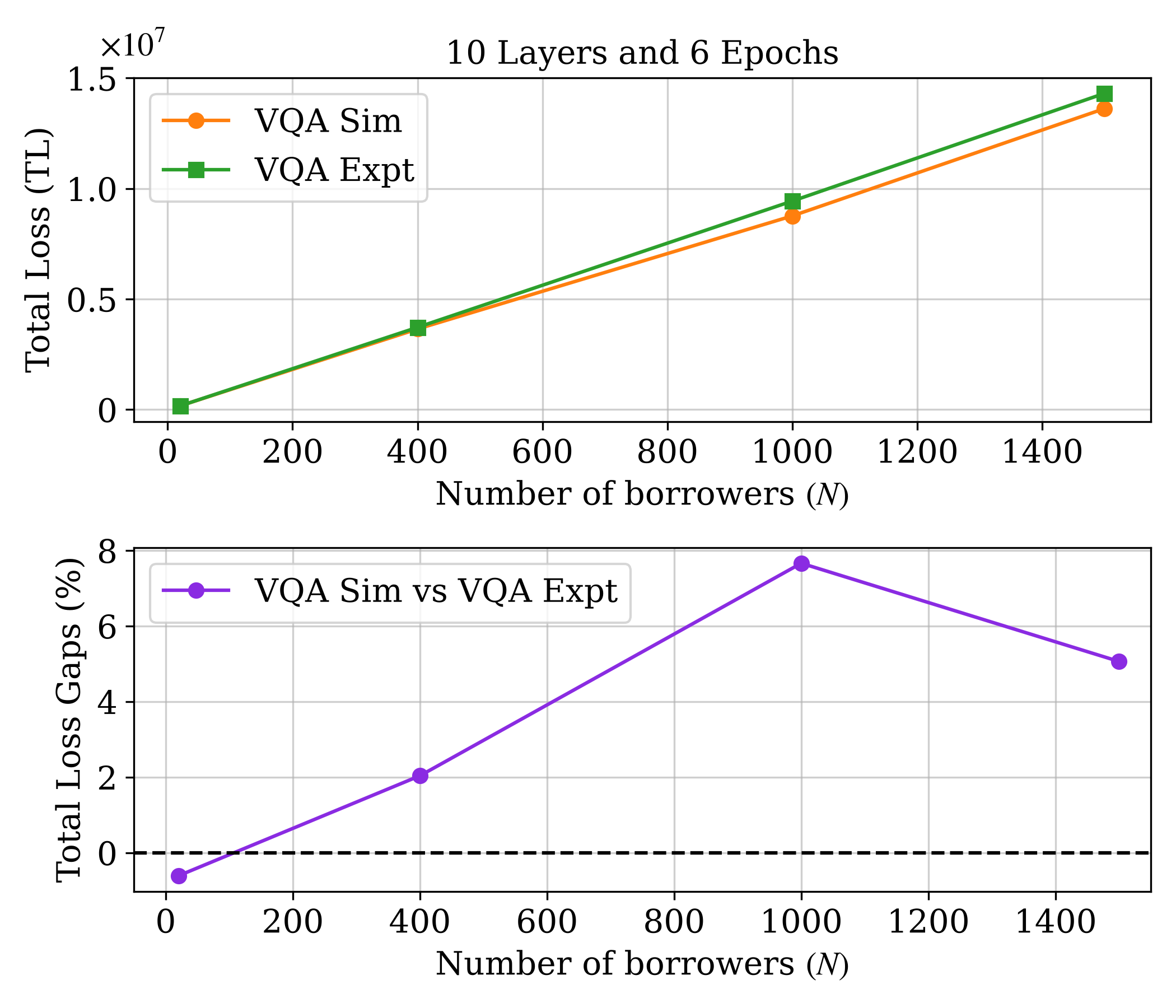}
    \caption{Comparison of VQA simulation and IBM quantum hardware results for four LPO instances with \(N=\{20,400,1000,1500\}\) and \(K=N/2\). Both approaches use the same ansatz architecture (10 layers), optimization schedule (6 epochs), hyperparameters, and initial parameters. The top panel shows the portfolio loss obtained from simulation and hardware execution, while the bottom panel shows the relative gap between the two solutions. Across all instances, the hardware results closely track the simulation results, with the relative gap remaining below \(10\%\).}
    \label{fig:VQA_sim_expt_N20_1500}
\end{figure}

Figure~\ref{fig:VQA_sim_expt_N20_1500} compares the solution quality obtained from VQA simulation and hardware execution for these four instances. The top panel shows the portfolio loss obtained by both approaches. Across all instances, the losses are closely matched, demonstrating strong agreement between simulation and hardware results. The bottom panel shows the relative gap between the two solutions, which remains below \(10\%\) for all instances.

The \(N=20\) instance corresponds to the dataset illustrated in Fig.~\ref{fig:Lit}, which shows the loss time series, and Fig.~\ref{fig:mu_sigma}, which shows the expected loss vector and covariance matrix. 
The corresponding portfolio solutions are visualized in Fig.~\ref{fig:N20_K10_VQA_BF}, while the portfolio-level (efficient-frontier-like) comparison is presented in Fig.~\ref{fig:N20_K10_VQA_BF_eff_front}.

%========================================================= 
\section{Discussion} 
\label{sec:discussion} 

We have presented a qubit-efficient framework for correlation-aware LPO based on variational quantum optimization and PCE. The portfolio model incorporates expected losses, \emph{loss variability}, and \emph{borrower-loss correlations}, and is formulated as a QCBO problem before being transformed into a QUBO representation suitable for VQAs. The use of PCE enables portfolio-selection problems involving hundreds to \emph{thousands of variables} to be represented using substantially \emph{fewer qubits} than conventional variable-to-qubit encodings.

The proposed framework was evaluated on a \emph{real-world} microfinance dataset comprising 2012 borrowers. Using PCE, we investigated optimization instances with up to 1500 portfolio-selection variables, exceeding the problem sizes considered in previous PCE-based equity portfolio optimization studies~\cite{soloviev2026}. For instances with up to 1000 variables, VQA solutions obtained from classical PQC simulations were benchmarked against Google OR-Tools under equal computational time budgets, yielding feasible portfolios with an average relative objective-value gap of approximately ${40\%}$. The framework was further implemented on IBM superconducting quantum hardware, where the experimentally obtained objective values remained within ${10\%}$ of the corresponding simulation results.

These results demonstrate the feasibility of applying PCE-based variational quantum optimization to \emph{large-scale} LPO under current quantum-hardware limitations. However, the present study does not establish a computational advantage over state-of-the-art classical optimization methods. Rather, it demonstrates that qubit-efficient encodings can substantially extend the size of portfolio optimization problems that can be represented and solved on contemporary quantum processors. More broadly, these results provide an empirical assessment of PCE as a resource-efficient representation for combinatorial optimization and contribute to ongoing efforts to understand the capabilities and limitations of quantum optimization in the NISQ regime.

%=========================================================
\section{Outlook} 
\label{sec:outlook}

Several important directions remain for future work. First, the current study derives PDs solely from credit scores. A more realistic framework should incorporate additional borrower-specific, behavioral, transactional, and macroeconomic factors to obtain more accurate risk estimates. Furthermore, we adopt simplified assumptions for EAD and LGD, treating EAD as the outstanding principal and assuming a zero recovery rate. In practice, accrued interest and non-zero recovery rates should also be considered to better reflect real lending scenarios.

Second, portfolio risk should be extended beyond expected and covariance-based losses to include tail-risk measures such as Conditional Value-at-Risk (CVaR), enabling optimization under extreme but plausible loss scenarios.

In addition, practical lending applications require operational and regulatory constraints that are not considered in the present formulation. Examples include portfolio-level risk-tolerance limits, exposure-concentration limits, and regulatory capital requirements under frameworks such as Basel III. Incorporating such constraints would yield a more realistic and practically relevant portfolio model.

Finally, quantum optimization alone may not always produce solutions that satisfy all practical constraints. Future work should therefore explore hybrid quantum-classical workflows in which quantum optimization is combined with classical post-processing or repair procedures to generate feasible portfolios while retaining the benefits of quantum search.

The scalability of the proposed Pauli-based representation also warrants further investigation. Although the encoding employed in this work allows the number of portfolio variables to grow exponentially with the number of qubits by exploiting the exponentially large Pauli operator space, the implications of this compression for optimization performance remain unclear. While the present study demonstrates successful optimization for problem sizes of up to 1500 borrowers, it remains unclear how the trainability of the associated VQA evolves as the number of encoded variables increases substantially beyond this scale.

Recent work on PCE suggests that higher-order Pauli-correlation representations can substantially mitigate the decay of gradient variances with problem size, thereby alleviating barren-plateau effects in certain settings~\cite{sciorilli2025}. Whether these advantages persist under more aggressive compression, however, remains an open question. Thus, while PCE provides an appealing approach for studying optimization problems with substantially more decision variables than those considered here, the relationship between representational and optimization scalability remains poorly understood. In particular, it is unclear how trainability, optimization performance, and solution quality evolve as the number of encoded variables grows well beyond the regimes explored in this work.

Highly compressed encodings also raise questions about classical simulability, as representational compactness alone does not necessarily translate into a computational advantage. Clarifying the interplay among compression, trainability, expressivity, and classical simulability will therefore be essential for assessing the ultimate scalability and computational potential of large-scale PCE-based quantum optimization frameworks. Collectively, these directions can help bridge the gap between proof-of-concept quantum portfolio optimization and real-world deployment.

%=========================================================
\section{ACKNOWLEDGMENTS} 
The authors thank Abhijit Sen, the IBM Quantum Research Lab, and Indranil Mitra for valuable discussions. The authors also acknowledge the use of large language models for language editing and polishing of the manuscript.

%=========================================================
\bibliographystyle{IEEEtran} 
\bibliography{ref}

\end{document}